\documentclass[sigconf]{acmart}
\AtBeginDocument{%
  }

\usepackage{xspace}
\usepackage{comment}

\setcopyright{acmlicensed}
\copyrightyear{2018}
\acmYear{2018}
\acmDOI{XXXXXXX.XXXXXXX}

\acmConference[Conference acronym 'XX]{Make sure to enter the correct
  conference title from your rights confirmation email}{June 03--05,
  2018}{Woodstock, NY}
\acmISBN{978-1-4503-XXXX-X/2018/06}

\begin{document}

\newcommand{\gnote}[1]{\textcolor{red}{[[[#1]]]}}
\newcommand{\totalSurveyResponses}{99\xspace}
\def\xxx{{\huge XXX}}


\title{Challenges and Enablers: Remote Work for People with Disabilities in Software Development Teams}
%\title{The Role of Remote Work in the Inclusion of People with Disabilities in Mixed-ability Software Development Teams}


%  (Zup Innovation & UFPA) <>
% Luciano Teran (UFPA) <Luciano.teran@icen.ufpa.br>
% Marcelle Mota (UFPA) <mpmota@ufpa.br>
% Cleidson de Souza (UFPA) <cdesouza@ufpa.br>
% Kiev Gama (Universidade Federal de Pernambuco) <kiev@cin.ufpe.br>
% Gustavo Pinto (Zup Innovation & UFPA) <gpinto@ufpa.br>

\author{Thayssa Rocha}
\orcid{1234-5678-9012}
\affiliation{%
  \institution{Zup Innovation \& UFPA}
  \city{Belém}
  \state{Pará}
  \country{Brazil}
}
\email{thayssa.rocha@icen.ufpa.br}

\author{Luciano Teran}
\affiliation{%
  \institution{UFPA}
  \city{Belém}
  \state{Pará}
  \country{Brazil}
}
\email{luciano.teran@icen.ufpa.br}

\author{Marcelle Mota}
\affiliation{%
  \institution{UFPA}
  \city{Belém}
  \state{Pará}
  \country{Brazil}
}
\email{mpmota@ufpa.br}

\author{Cleidson de Souza}
\affiliation{%
  \institution{UFPA}
  \city{Belém}
  \state{Pará}
  \country{Brazil}
}
\email{cdesouza@ufpa.br}

\author{Kiev Gama}
\affiliation{%
  \institution{UFPE}
  \city{Recife}
  \state{Pernambuco}
  \country{Brazil}
}
\email{kiev@cin.ufpe.br}

\author{Gustavo Pinto}
\affiliation{%
  \institution{UFPA \& Zup Innovation}
  \city{Belém}
  \state{Pará}
  \country{Brazil}
}
\email{gpinto@ufpa.br}

\renewcommand{\shortauthors}{Trovato et al.}

\begin{abstract}
The increasing adoption of remote and hybrid work modalities in the technology sector has brought new opportunities and challenges for the inclusion of people with disabilities (PWD) in software development teams (SDT). This study investigates how remote work affects PWDs' experience in mixed-ability SDT, focusing on the unique challenges and strategies that emerge in remote environments. We conducted an online survey with \totalSurveyResponses valid responses, encompassing PWD, their leaders, and teammates, to capture sociotechnical aspects of their experiences with remote collaboration. To deepen our understanding, we carried out 14 structured interviews with software developers who self-identified as having disabilities (six autistic individuals, six with physical disabilities, and two who are d/Deaf). Our analysis combines quantitative data with qualitative coding of open-ended survey responses and interview transcripts. The results reveal that, despite the barriers faced by team members with disabilities, their teammates and leaders have a limited perception of the daily challenges involved in sustaining collaborative remote work. These findings highlight opportunities for improvement in accessibility tools, communication strategies, and adaptive management approaches.\end{abstract}



%\keywords{Do, Not, Use, This, Code, Put, the, Correct, Terms, for,  Your, Paper}



\received{20 February 2007}
\received[revised]{12 March 2009}
\received[accepted]{5 June 2009}

\maketitle

\section{Introduction}

Although many countries have established cooperation agreements over the past decade to promote the inclusion of people with disabilities (PWD) in the workplace \cite{UN_CRPD_2006}, achieving effective inclusion for PWD in this area continues to be a global challenge \cite{WHO2022GlobalReport}.
International Labour Organization (ILO) highlights persistent barriers faced by PWD in the workplace , including limited access to adequate resources, lack of organizational accommodations, and societal stigma that affects hiring, career advancement, and day-to-day interactions~\cite{ILO_InclusiveEmployment_2016, ILO_WSPR_MENA_2021, ILO_Workers_Disabilities_Jordan_2022}. These barriers are particularly significant in knowledge-intensive industries, such as technology, where collaboration, problem-solving, decision-making, and continuous learning are central to productivity and success~\cite{ILO_ESG_2024}. 

Software development teams rely heavily on constant communication and activities coordination~\cite{Curtis-1988}, often mediated through digital platforms~\cite{Plonka2015}. When accessibility is not considered, PWD can face exclusion from critical interactions, limiting both their individual contributions and the overall performance of the team. As remote and hybrid work models become increasingly common in the technology sector, the need to understand and address these issues has never been more pressing.

Despite the growing prevalence of distributed software development teams, little is known about how PWD experience remote work within mixed-ability teams—teams composed of both disabled and non-disabled professionals. Prior research on remote software development has primarily focused on general collaboration challenges, such as communication breakdowns, coordination difficulties, and the integration of tools and practices across distributed teams~\cite{bird2011sociotechnical,ford2021tale}. However, these studies often overlook the additional complexities introduced by accessibility needs and disability-related barriers. Consequently, there is a lack of empirical evidence on how PWD navigate these remote environments, what specific challenges they encounter, and which factors foster their inclusion and active participation. This gap leaves both researchers and practitioners without clear guidance on how to design effective, inclusive practices for distributed teams. To approach this matter, we raised three research questions: \textbf{RQ1:} How do disabled and non-disabled team members evaluate remote work in mixed-ability software development teams (SDT)? \textbf{RQ2:} What are the main challenges faced by PWD in remote mixed-ability SDT? and \textbf{RQ3:} What aspects promote a positive experience of remote work for PWD in SDT?

To address these questions, we conducted a comprehensive explanatory sequential study. ``Sequential'' because the two strands are done in sequence (one after the other) and ``Explanatory'' because the qualitative phase is used to help explain or elaborate on the quantitative results. The study was conducted in a large Brazilian consulting company, and we combined quantitative and qualitative approaches~\cite{creswell2014research}
to capture both the breadth and depth of experiences. First, we administered an online survey and collected 99 responses from participants of three groups: PWD, Leaders, and non-PWD teammates, to explore general perceptions of remote work and identify patterns related to accessibility and inclusion. Building on these results, we conducted 14 semi-structured interviews with PWD developers to gain deeper insights into the challenges and positive factors that shape their experiences in remote software teams. %By integrating findings from both data sources, 
This research provides an empirical understanding of how PWD and non-PWD collaborate in distributed settings, the barriers that hinder effective participation, and the practices and strategies that can foster more equitable and inclusive software development environments. In summary, our contribution is an in-depth understanding of the challenges and enablers associated with having PWD as members of software development teams.

%\gnote{falta escrever aqui os principais achados}

%The results reveal that, PWD face different barriers including \textcolor{red}{A, B, C and D}. Despite these barriers, their teammates and leaders have a limited perception of the daily challenges involved in sustaining collaborative remote work. For instance, \textcolor{red}{bla bla bla}

%These findings highlight opportunities for improvement in accessibility tools, communication strategies, and management approaches.

From the survey and interviews, five key findings emerged that address our research questions:  
\begin{itemize}
    \item Remote work was broadly evaluated as positive across all groups (PWD, leaders, and teammates), with higher satisfaction rates in remote-first arrangements compared to hybrid ones.  
    \item Communication in remote teams is generally perceived as inclusive and effective, but only PWD reported neutral or negative experiences, pointing to subtle accessibility barriers not perceived by others.  
    \item Coordination was mostly well-evaluated, yet PWD were more likely to report difficulties in task clarity and workflow transitions, highlighting gaps in transparency and documentation.  
    \item Cooperation was rated highly by all groups, though a minority of PWD still reported barriers, suggesting that some challenges remain invisible to non-disabled peers and leaders.  
    \item Positive experiences were strongly associated with supportive team culture, clear and respectful communication, and the personal benefits of remote work (flexibility, autonomy, and quality of life), which were especially valued by PWD.  
\end{itemize}

% The increasing adoption of remote and hybrid work modalities in the technology sector has brought new opportunities and challenges for the inclusion of people with disabilities (PWD) in software development teams (SDT). This study investigates how remote work influences PWD's inclusion in mixed-ability SDT, focusing on the unique challenges and strategies that emerge in remote environments.

% General Objective: Understand remote work influences the participation of PWD in mixed-ability SDT.

% Specific Objectives:
% \begin{itemize}
%     \item Identify the main challenges and benefits faced by PWD in mixed-ability remote SDT
%     \item Analise the non-disabled pairs perception on these challenges and benefits and their involvement in turning the remote environment more inclusive
% \end{itemize}

% Research Questions:
% \begin{itemize}
%     \item SQ1 - How do disabled and non-disabled pairs evaluate remote work in mixed-ability Software Development Teams (SDT)?
%     \item SQ2 - What are the main challenges faced by PWD in remote mixed-ability SDT?
%     \item SQ3 - What aspects promote a positive experience of remote work for PWD in SDT?
%   % ACHO QUE VOU RETIRAR ISSO E DEIXAR COMO CONCLUSÕES  \item SQ4 - What are the benefits of remote work to PWD in SDT?
% \end{itemize}

\section{Background}

According to Santos et al.~\cite{ronniegrounded} there are two particular kinds of remote work: \textit{remote-first team} and \textit{Hybrid-team}.

\begin{itemize}
    \item \textit{A remote-first team}. The default work arrangement is that each team member works from their own location, such as a home office, coffee shop, or co-working space, and a centralized office may or may not exist.
    \item \textit{A Hybrid-team} On any given day, some members work from a shared office while others work remotely, with configurations varying depending on schedules, roles, or personal preferences.
\end{itemize}

In this work, the authors focused on coordination challenges in remote-first/hybrid software teams, which are mainly influenced by factors like trust, cohesion, communication breakdowns, parenting, and bricolage of tools. Remote-first coordination bring structural issues (communication, trust, cohesion) that can also affect accessibility and inclusion of persons with disabilities in software teams.

% Remote work Concepts (remote first e hibrid)=> \cite{ronniegrounded}

% Falar sobre a falta de papers: \cite{thayssaESEM}

A recent paper~\cite{RochaSTM24}, the authors conducted a systematic literature review (SLR) to understand the human and technical perspectives in the context of PWD in software development. Out of the 572 papers identified from 2013 to 2023, only 39 were selected to be analyzed in depth. The main challenges reported occur in the context of teamwork dynamics in mixed-ability teams and tool accessibility in daily work activities. In general, this SLR suggests that studying PWD as members of software development teams is barely explored.

\subsection{Related Works}
Some researchers have addressed PWD's professional remote experiences; however, most research focuses on one aspect of work collaboration or considers general field areas, not focusing on software development's unique characteristics.

Santos and Ralph \cite{santos2022agrountheory} conducted a year-long observational study during the COVID-19 pandemic with workers at a South American company to identify the challenges faced by software development professionals in remote and hybrid environments. Applying Constructivist Grounded Theory to the data, the researchers presented four recommendations to help managers and members of remote and hybrid teams recognize coordination issues affecting software teams in the post-pandemic era. Building on this research and the challenges faced by software development teams, we sought to investigate factors that impact the productivity of professionals with disabilities in hybrid and remote software environments.

Akter et al.~\cite{Akter2023Meet} conducted a set of interviews with 18 meeting facilitators for people with blindness and low vision (BLV) regarding their practical experiences and challenges in using videoconferencing tools. The researchers found that facilitators needed to perform preparatory activities and develop skills to overcome accessibility issues to conduct remote meetings effectively. The facilitators also sought ways to navigate these tools without affecting their productivity. The contributions highlighted the need to improve the accessibility and usability of meeting tools, promote accessible meeting practices and norms, and support access at an organizational level to enhance the remote workplace experience for BLV professionals.

Cha et al. \cite{cha2024doyouwhat} interviewed 26 BLV software professionals in a semi-structured format about the accessibility of meeting tools. Based on four themes related to in-person and remote software development meetings, the researchers presented a set of design implications for the inclusion of BLV professionals in remote environments: technical and policy-driven solutions for accessible meetings and low vision research. Related issues were also reported in Tang's~\cite{tang2021understanding} research about remote work involved individuals with various disabilities: blind or low vision, deaf or hard of hearing, neurodiverse, people with limited mobility/dexterity, and chronic health issues. Roughly one third of participants were IT-related professionals. The findings illustrate that technology readiness is limited for disabled individuals. For instance, telework tools can misrepresent or marginalize disabled users. Design choices in video-call interfaces demote people who turn cameras off (e.g., blind users) to static name tiles, and active-speaker layouts can exclude Deaf participants speaking via interpreters. Screen sharing and collaborative editors may also reveal disability-related cues, prompting some users to avoid these features.

In a study performed during the COVID-19 pandemic, Das and colleagues~\cite{das2021towards} highlighted that working from home gives neurodivergent professionals more flexibility, but it also shifts heavy cognitive and emotional labor onto them~\cite{das2021towards}.  
Liebel et al\cite{liebel2024challenges} studied difficulties of ADHD developers and flexible work arrangements such as remote work was one of the strategies recommended.

Despite the importance of investigating accessible professional tools for BLV, as in the studies by Akter et al.~\cite{Akter2023Meet}, Cha et al. \cite{cha2024doyouwhat} and Tang~\cite{tang2021understanding}, our research stands out for its combination of questionnaires and interviews, involving people with and without disabilities, to identify challenges and broader insights for building more inclusive software development teams, aiming to find positive aspects generated by stakeholders, processes and also tools that act and collaborate to meet the conditions of diverse PWD professionals in their workplaces.

%Cha et al. \cite{cha2024doyouwhat}

% In a more contextualized approach, Cha's research investigate the accessibility of software development team meetings for blind and low vision professionals. Results revealed systemic barriers in remote and in-person modalities. Participants report extensive access labor, challenges with screen sharing and visual tools, and the emotional toll of disability disclosure.

%\cite{100pressured}.

\section{Research Questions}

This study aims to explore the experiences of people with disabilities working in remote and hybrid software development teams. Based on the gaps identified in the literature and the growing need to understand inclusion in distributed software engineering environments, we formulated three research questions. Each research question addresses a specific dimension of this phenomenon, from the general perception of remote work to the identification of barriers and enablers for inclusion, from various perspectives.

%\subsection*{RQ1: How do disabled and non-disabled people evaluate remote work in mixed-ability software development teams?}

\textbf{RQ1: How do disabled and non-disabled people evaluate remote work in mixed-ability software development teams?}

This research question seeks to understand how both disabled and non-disabled team members perceive remote work within mixed-ability teams. Remote work has become a dominant model in the technology sector, especially after the COVID-19 pandemic, bringing changes to communication, collaboration, and productivity~\cite{ford2021tale}. However, while prior studies have investigated these issues broadly, they rarely account for how accessibility needs influence these experiences. Understanding how different groups evaluate remote work is essential for identifying whether PWD face unique challenges or whether their experiences align with those of their non-disabled peers. 

%\subsection*{RQ2: What are the main challenges faced by PWD in remote mixed-ability SDT?}
\textbf{RQ2: What are the main challenges faced by PWD in remote mixed-ability SDT?}

The second research question focuses on uncovering the specific challenges encountered by PWD in remote or hybrid SDTs. Existing literature has highlighted common problems in distributed teams, such as coordination difficulties, communication breakdowns, and tool integration issues \cite{treude2012programming,ford2021tale, machado2021gendered}, but little is known about how these challenges manifest for PWD. By identifying and categorizing the barriers faced by PWD, this research question aims to provide a deeper understanding of how disability intersects with distributed work practices. Addressing this gap is critical for organizations seeking to remove systemic barriers, ensure accessibility, and create equitable working conditions. Furthermore, understanding these challenges contributes to advancing research on diversity and inclusion in software engineering by expanding current knowledge beyond demographic representation to the lived experiences of marginalized groups.

%\subsection*{RQ3: What aspects promote a positive experience of remote work for PWD in SDT?}
\textbf{RQ3: What aspects promote a positive experience of remote work for PWD in SDT?}

While RQ2 focuses on barriers and challenges, RQ3 explores the factors that enable positive and inclusive experiences for PWD in remote software development contexts. Identifying enablers is equally important as it shifts the focus from problems to solutions. By examining what works well, we can uncover practices, tools, and strategies that foster meaningful participation and collaboration for PWD. This includes understanding how team culture, leadership practices, and technology choices can contribute to accessibility and belonging.

\section{Methods}

To investigate the experiences and challenges faced by programmers with disabilities in software development teams, we adopted a mixed-methods approach that combined both quantitative and qualitative research methods. Mixed-method designs are widely recommended in empirical software engineering research because they provide complementary strengths: quantitative data enables generalization across a population, while qualitative data offers rich, contextualized insights into complex socio-technical phenomena~\cite{stol2016grounded,runeson2012case}.

Specifically, we adopted an explanatory sequential design 
%\textcolor{red}{Creswell} 
in which we started with a quantitative approach (a survey) followed by a qualitative approach (interviews). The survey data established broad patterns and prevalence rates of accessibility issues, while the interview data offered rich narratives that contextualized these patterns and revealed underlying socio-technical mechanisms.
%In this study, we leverage surveys and semi-structured interviews. 
The combination of surveys and semi-structured interviews enabled data triangulation, which strengthens the validity of findings by providing both breadth and depth of understanding~\cite{runeson2012case}. %Specifically, the survey data established broad patterns and prevalence rates of accessibility issues, while the interview data offered rich narratives that contextualized these patterns and revealed underlying socio-technical mechanisms. 

This study is part of a wider research project about PWD as members of software development teams. It was previously reviewed and approved by the Ethics Committee in our university. 

\subsection{Survey}
\label{sec:survey}
Surveys are a well-established method in software engineering research for capturing a wide range of perceptions, practices, and attitudes at scale~\cite{wright2017survey}. 
An \textit{online survey} was first distributed to a broad group of stakeholders involved in software development within a large Brazilian consulting company. It was conducted between July and August of August 2025 and remained open for two windows of 15 days. It was disseminated to employees across different roles in the company to ensure a diversity of perspectives and experiences. To guarantee the integrity of the data and prevent duplicate submissions, the survey was not anonymous. This approach allowed us to ensure that each participant could respond only once, thereby increasing the reliability of the collected data. In total, we collected \totalSurveyResponses valid responses, consisting of 63 individuals with disabilities, 17 team leaders, and 19 teammates without disabilities. %More about the participant profiles in Section \xxx.

The survey was %contained a total of \xxx questions, 
divided into five sections to capture both individual and team-level perspectives. The \textit{first} section gathered demographic data, including age, gender, role in the development team, and whether participants identified as a person with a disability. The \textit{second} and \textit{third} sections branched depending on disability status: participants with disabilities were asked about the type of disability, use of assistive technologies, and leadership roles over other PWD, while participants without disabilities reported on their interactions with colleagues with disabilities. The \textit{fourth} section focused on characterizing remote work practices, covering team size, work format (e.g., fully remote or hybrid), duration in the current format, and satisfaction levels. Finally, the \textit{fifth} section, answered by all respondents explored three key aspects of teamwork in remote settings: communication, coordination, and cooperation, through questions about the tools used for these aspects, perceived quality (rated on a five-point Likert scale), and open responses to explain their evaluations. This aspect is \textit{not} discussed in this paper. Some representative questions are listed below to illustrate the structure and scope of the instrument.

\vspace{0.2cm}
\noindent\textbf{(1) Demographics}
\begin{itemize}
    \item What is your age range? [\textit{Options: Under 18 / 18--24 / 25--34 / 35--44 / 45--54 / 55--64 / 65 or older / Prefer not to answer / Other (free text input)}]
    \item Which gender do you identify with? [\textit{Options: Cisgender Woman / Cisgender Man / Trans Woman / Trans Man / Non-Binary Person / Prefer not to answer / Other (free text input)}]
    \item What is your current role in the development team?
    \item Do you identify as a person with a disability?
\end{itemize}

\noindent\textbf{(2) For Participants with Disabilities}
\begin{itemize}
    \item You identified yourself as a person with a disability. Which group(s) do you belong to?
    \item In your daily work, do you use any assistive technologies, such as screen readers, adapted mouse, sign language translator, automatic caption generator, or others? If yes, please specify.
    \item Do you act as a functional leader for other people with disabilities? [\textit{Yes/No}]
\end{itemize}

\noindent\textbf{(3) For Participants without Disabilities}
\begin{itemize}
    \item Do you interact with one or more people with disabilities in your daily work? [\textit{Yes/No}]
\end{itemize}

\noindent\textbf{(4) Remote Work Characterization}
\begin{itemize}
    \item How many people are part of your squad/team? [\textit{open-ended response}]
    \item What is your current work format? [\textit{Options: My team and I are 100\% remote (we never need to go to the office) / I am 100\% remote, but other team members need to go to the office one or more times per week / My team and I need to go to the office one or more times per week / Other (free text input)}]
    \item How long have you been working in this format? [\textit{open-ended response}]
    \item How do you rate your satisfaction with this work format? [\textit{Scale from 1 to 5, where 1 = very dissatisfied and 5 = very satisfied}]
\end{itemize}

\noindent\textbf{(5) 3 C's of Remote Work Aspects (Communication / Coordination / Cooperation) - For PWD} %\textcolor{red}{citar artigo em inglês do 3C aqui.}}
\begin{itemize}
    \item Which \textit{communication} tools do you use to interact with your coworkers? [\textit{Email / Group chats / Direct chats / Phone calls / Video calls / WhatsApp - Telegram / Other}]
    
    \item How do you evaluate \textit{communication} with your coworkers? Use a scale from 1 to 5, where 1 = Very poor and 5 = Very good. [\textit{1/2/3/4/5}]
    
    \item What led you to evaluate \textit{communication} this way? [\textit{open-ended}]
    
    \item For \textit{coordinating}\footnote{Coordinating activities include task such as defining or tracking who is responsible for what, identifying dependencies, start and end of tasks, and handoffs between team members' tasks.} activities with your coworkers, which tools do you use?
    Select all options that apply. [\textit{Email / Shared spreadsheet / Management tool (e.g., Trello, Jira, proprietary tool, etc.) / Other}]
    
    \item How do you evaluate activity \textit{coordination} with your coworkers? Use a scale from 1 to 5, where 1 = Very poor and 5 = Very good. [\textit{1/2/3/4/5}]
    
    \item What led you to evaluate activity \textit{coordination} this way? [\textit{open-ended}]
    
    \item Which \textit{cooperation}\footnote{Cooperation include activities such as working together (synchronously or asynchronously) with one or more colleagues on a specific task to produce a single, shared outcome.} resources do you use with your coworkers?  
    Select all options that apply. [\textit{Joint editing of documents (code, text, spreadsheets, diagrams, etc.) / Screen sharing / Video call / Voice call / Group chat / Other}]
    
    \item How do you evaluate \textit{cooperation} with your coworkers? Use a scale from 1 to 5, where 1 = Very poor and 5 = Very good. [\textit{1/2/3/4/5}]
    
    \item What led you to evaluate \textit{cooperation} this way? [\textit{open-ended}]
\end{itemize}

Before deploying the final version of the survey, we conducted a pilot study with two participants with a disability and one leader without disability. The primary goal of this pilot was to ensure that the questions were clearly formulated, unambiguous, and accessible, as well as to verify that participants had sufficient time to read, reflect, and provide thoughtful responses. Conducting a pilot survey is a well-established practice in empirical research, as it helps to identify potential issues related to question wording, structure, and usability before launching the full study~\cite{wright2017survey}. Based on the feedback received from the pilot participants, we made a few minor adjustments to improve the overall clarity and accessibility of the survey. For instance, we refined the descriptive text accompanying the Likert scale questions to make them easier to understand and more compatible with screen readers used by participants with visual impairments. In other words, piloting a survey is particularly important when studying PWD.

The survey instrument was carefully designed to capture participants' perceptions of collaboration practices, accessibility barriers, and inclusion within their teams. By leveraging this large-scale data collection approach, we were able to gather %statistically significant 
evidence that provided a foundational understanding of the prevalence and distribution of accessibility challenges across different roles in software development teams.

\subsection{Interviews}

%To complement the breadth of the survey data and gain a deeper understanding of individual experiences, we conducted a series of \textit{semi-structured interviews}. Interviews are particularly valuable in software engineering studies because they allow researchers to explore participants' perspectives in detail, uncovering nuanced insights that are difficult to capture through predefined survey questions \cite{seaman1999qualitative}. Two pilot interviews were conducted initially to refine the interview protocol and ensure the clarity and relevance of the questions. The final set of interviews included 14 programmers with disabilities: six autistic individuals, six with physical disabilities, and two who were deaf. This sampling strategy ensured representation of diverse disability experiences and allowed us to explore a range of accessibility and inclusion challenges faced by software engineers in different contexts.

To complement the breadth of the survey data and gain a deeper understanding of participants' individual experiences, we conducted a series of 14 \textit{semi-structured interviews}. While the survey provided a broad overview of patterns and trends regarding accessibility, remote work practices, and collaboration challenges, the interviews allowed us to explore these topics in greater depth, uncovering nuanced insights that are difficult to capture through predefined, close-ended survey questions \cite{seaman1999qualitative}. The interview protocol was designed to build upon key themes identified in the survey, such as communication, coordination, and cooperation in remote settings, as well as team dynamics, leadership, and organizational support for inclusion. In some points the interviews were conducted using the participants' survey answers as an input to allow interviewees reflect and expand their answers.

Two pilot interviews were conducted: one with a blind and other with a deaf programmer. The goal was to refine the interview protocol ensuring the questions' clarity, accessibility, and relevance to participants, test the logistics (tool recording, etc), and evaluate whether the answers are producing the depth and richness required. 
Based on the feedback received, minor adjustments were made. Some questions were rewritten to be more clear, and a few questions were merged, since they approached the same issue. Some representative questions are listed below.

%\gnote{tem que remover algumas perguntas aqui}

\vspace{0.2cm}
\noindent\textbf{(1) Characterization}  
\begin{itemize}
    \item How long have you been working in this role? 
    \item What is your seniority level?
    %\item You self-identified as \xxx (information obtained from the survey). Could you share a bit about your diagnosis?
    \item (If neurodivergent) Are you on the autism spectrum? What is your level of support?
    \item (If visually impaired) What percentage of visual loss do you have?
\end{itemize}

\noindent\textbf{(2) Team and Leadership}  
\begin{itemize}
    \item Tell me a bit about your team.
    \item Are there other PWD in your team?
    \item Do your colleagues know about your diagnosis?
    \item Is your direct manager a person with a disability?
    \item Does your manager know about your diagnosis?
\end{itemize}

\noindent\textbf{(3) Education and Training}  
\begin{itemize}
    \item Tell me a bit about your career path.
    \item Have you been part of any professional training program?
    \item Tell me about your formal education.
    \item What are the main challenges you have faced in accessing professional education and training?
    \item How do you overcome these challenges?
\end{itemize}

\vspace{0.2cm}
\noindent\textbf{(4) Work Dynamics}  
\begin{itemize}
    \item Tell me about your experience working in your current role.
    %\item You rated your satisfaction with this work modality as \xxx (information obtained from the survey). Why?
    \item Have you worked in-person before? What were the pros and cons compared to your current work format?
    \item How does remote work affect your employment opportunities?
    \item How would you describe your experience as a member of the development team?
    \item Do you feel included and part of the team?
    \item What factors contribute to successful teamwork?
    \item What could be improved to enhance your experience?
    \item Have you or your team adopted specific practices, tools, or strategies to deal with day-to-day challenges? Could you share an example?
    \item (If applicable) Does the hybrid format of some members influence your perception of teamwork? How?
\end{itemize}

\vspace{0.2cm}
\noindent\textbf{(5) Collaboration Aspects}  

\noindent\textit{(5.1) Shared Understanding}  
\begin{itemize}
    \item How is information and knowledge (technical and business) shared within your team?
    \item Are there factors that facilitate or hinder mutual understanding between you and other team members?
    \item How about the sharing of information between PWD and non-PWD team members?
    \item How much of your work depends on interaction with colleagues?
    \item Do you feel you have all the knowledge you need to perform your tasks?
    \item How do you acquire this knowledge?
    \item When you first joined the team, how did you learn about the workflows, tools, meetings, and practices?
    \item Do you feel you clearly understand the tasks, objectives, and collective decisions?
    \item Do you feel your colleagues understand your ideas and information needs? What helps or hinders mutual understanding?
    \item Is this your first time working with this team?
    \item How would it be for you to change teams and have to rebuild this shared understanding?
\end{itemize}

\noindent\textit{(5.2) Work Coupling}  
\begin{itemize}
    \item Do you usually receive isolated tasks, or are your tasks well-integrated with the team's sprint activities?
    \item Would you like to have more interaction, or do you think the current level of integration is ideal?
    \item How does this affect your sense of belonging to the team?
    \item How do you and your team track the progress of each other's tasks (e.g., meetings, tools)?
    \item How do you evaluate your participation and understanding during these events?
    \item Which practices or tools help (or could help) with understanding and integrating team activities?
    \item Do you face difficulties in establishing agreements with other team members to perform dependent or joint activities? How do you overcome these challenges?
\end{itemize}

The final set of interviews included 14 programmers with disabilities: six autistic individuals, six with physical disabilities, and two who were deaf. %More information about them is presented in Table ???.

%\gnote{na perspectiva de usar mais espaço, podemos criar uma tabela com os demographics dos entrevistados. thay: sim, eu gosto. eu explico tb que os entrevistados foram selecionados dauqlees que se dispuseram no survey e que eu escolhi vomecar por esses por causa da representatividade dentre os programadores? todos os 14 sao programadores. apesar de termos recebido outros papeis nao entrevistei outros papeis(ainda)} 

Our purposeful sampling strategy ensured representation across different types of disabilities, enabling us to explore a wide range of accessibility and inclusion challenges experienced by software engineers in diverse remote and hybrid work contexts. In total, the interviews were conducted between the beginning of July and the end of August, 2025. On average, each interview lasted approximately one hour.

\subsection{Data Analysis}

We employed diverse data analysis methods, according to the data collected. To analyze the survey described in Section \ref{sec:survey}, we used descriptive statistics to provide a concise summary of the primary information. 

For interviews, all interviews were transcribed using AI-powered transcription tools and subsequently manually reviewed for accuracy. In cases where the participant had speech impairments or where the AI-generated transcription was unclear, the transcription was performed entirely by hand to ensure the fidelity and reliability of the data.

For the open-ended survey questions and interview transcripts, we applied open coding techniques following a thematic analysis approach~\cite{braun2006using,cruzes2011recommended}. Initially, the first author systematically reviewed the data, generating a preliminary set of codes by identifying recurring keywords, patterns, and relevant concepts. As the analysis progressed, this initial coding scheme was iteratively refined through discussions among the research team, ensuring that multiple perspectives were considered and reducing potential researcher bias.  Using this refined coding scheme, we collaboratively analyzed the data to group similar codes into higher-order themes that represented broader concepts and relationships. Throughout this process, we revisited and adjusted both the codes and themes to ensure alignment with the underlying data. Final themes and their nomenclature were validated through consensus meetings among all authors, resolving disagreements through discussion. 

The goal of this analysis was to provide a deep understanding of the challenges and enablers experienced by PWD when participating in hybrid and remote software development teams.

% \textcolor{red}{Thayssa - ajustar
% For the open-ended questions from the survey and interview transcripts, open coding techniques were utilized to classify the answers [citar thematic analysis aqui]. Initially, the first author took on the task of creating an initial list of codes and identifying keywords or relevant codes in the interview data. As the process advanced, the list was reviewed and refined by other authors. Based on this initial list of codes, we conducted repeated collaborative analyses
% to identify similarities between the codes to develop higher-order codes or categories. During the process, we also recognized the
% need to refine the identified categories. Finally, we conducted a thorough review of the categories and made modifications to their
% nomenclature when necessary to ensure that they represented the information found in the interview data more clearly. Any conflicts were resolved through discussion and consensus meetings.}

% The aim of this analysis was to understand the  challenges and enablers faced by PWD when working as software development team members in hybrid and remote contexts. 

\section{Results}

\subsection{Satisfaction with Remote Work}

To answer\textbf{ RQ1 - How do disabled and non-disabled team members evaluate remote work in mixed-ability software development teams}?, we analyzed quantitative and qualitatively the results from online survey.

\begin{figure}[h]
  \centering
  \includegraphics[width=\linewidth]{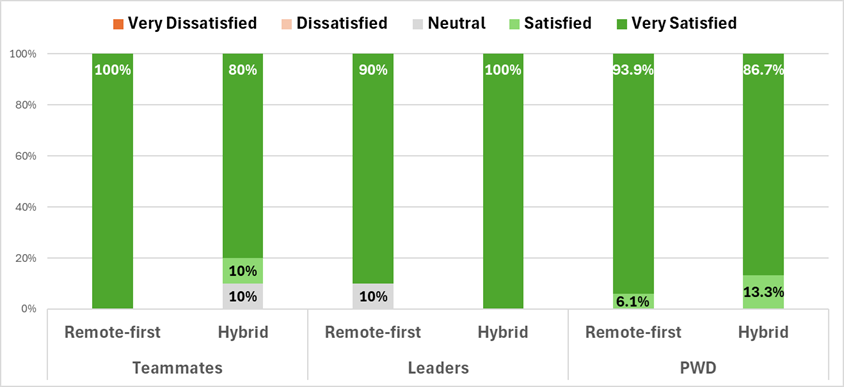}
  \caption{Satisfaction with remote work}
  \Description{Bar chart comparing satisfaction levels across three groups—Teammates, Leaders, and Persons With Disabilities (PWD)—based on two work arrangements: Remote-first and Hybrid. Each bar is color-coded to represent satisfaction levels: Very Dissatisfied (dark orange), Dissatisfied (light orange), Neutral (gray), Satisfied (light green), and Very Satisfied (dark green).}
  \label{fig:satisfaction}
\end{figure}

Figure~\ref{fig:satisfaction} shows that when comparing quantitative results on satisfaction among PWD, their teammates, and leaders, we can notice that remote work is perceived very positively across all groups, with minimal dissatisfaction. Teammates who work with remote-first modality expressed as 100\% Very Satisfied, while those who work in hybrid teams expressed 80\% Very Satisfied, 10\% Satisfied, and 10\% Neutral. 90\% of Leaders in remote-first work ranked their experience as Very Satisfied, and 10\% as Neutral, while in hybrid work, they ranked their experience as 100\% Very Satisfied. Finally, PWD working in remote-first presented a higher rate of satisfaction (6.1\% Satisfied and 93.9\% Very Satisfied) than those in hybrid arrangements (13.3\% Satisfied and 86.7\% Very Satisfied)
The chart highlights higher satisfaction levels in remote-first arrangements across all groups, with hybrid arrangements showing slightly more neutral or dissatisfied responses.

However, the positive results are also common in remote-first and hybrid arrangements, showing a slight variation from other evaluations (Satisfied and Neutral), especially in hybrid modality.
This convergence suggests that remote work is broadly perceived as a positive modality, not only by those in leadership positions but also by team members with and without disabilities. %This may indicate that remote work environments meet the diverse needs of these groups to a significant extent, potentially offering greater flexibility, autonomy, and accessibility.

%In the following paragraphs, we will detail the results of each group: Leaders, PWD and PWD teammates.

% \subsubsection{Leaders' Perspective}

Based on the Leaders' qualitative survey answers, there is a clear sense of satisfaction with remote work in mixed-ability teams. Leaders consistently report that communication has not been a challenge (see Figure \ref{fig:communication}), with most interactions flowing smoothly and effectively, even when working with team members with hearing or visual impairments. This quantitative result suggests that the use of interpreters, descriptive communication, and accessible digital tools has enabled clear communication with Leaders.

The PWD group also reported a high level of satisfaction with remote work in mixed-ability software development teams (see Figure \ref{fig:satisfaction}). The open answers of the survey suggest that 
communication is generally clear, respectful, and inclusive, with colleagues making efforts to ensure everyone understands the demands and has access to meeting notes for later review. The flexibility of remote work is especially valued, as it allows for better work-life balance, saving time previously lost in commuting and enabling more time for health, study, and family. Teams are described as engaged, supportive, and collaborative, with multiple channels for communication—such as chat, video calls, and written messages—accommodating different preferences and needs. There is a strong sense of partnership, with team members always willing to help, answer questions, and adapt communication styles when necessary. While some challenges remain, such as occasional difficulties with asynchronous communication or the accessibility of specific tools, the overall experience is positive, with people feeling included, respected, and able to contribute effectively to their teams.

\subsubsection{Communication Results}
\begin{figure}[h]
  \centering
  \includegraphics[width=\linewidth]{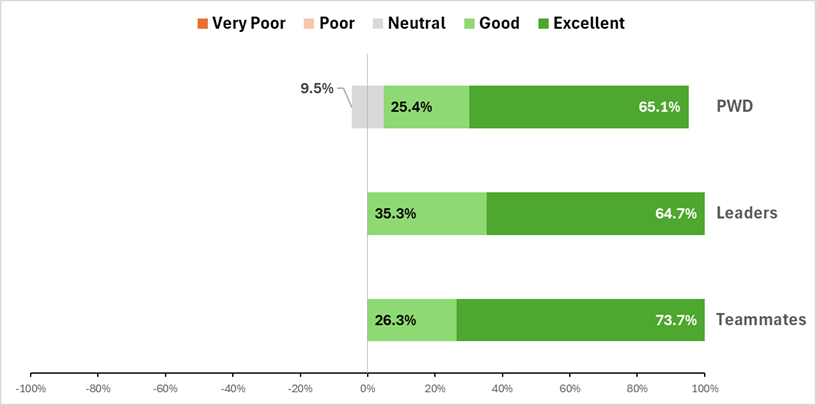}
  \caption{Evaluation of remote \textbf{communication}.}
  \Description{Horizontal bar chart comparing perceptions about remote communication among PWD, Leaders, and Teammates. Each group is represented by a segmented horizontal bar, color-coded by rating categories: Very Poor (dark orange), Poor (light orange), Neutral (gray), Good (light green), and Excellent (dark green).}
  \label{fig:communication}
\end{figure}

Quantitative data reveals nuances in how different groups in remote, mixed-ability teams experience \textbf{communication}. Figure~\ref{fig:communication} shows that among PWD, 9.5\% rated their communication experience as ``Neutral'', 25.4\% as ``Good'' and 65.1\% as ``Excellent"'. For leaders 35.3\% rated ``Good'', 64.7\% rated ``Excellent''. 26.3\% of teammates rated ``Good'' and  73.7\% as "Excellent".

The relatively high positive assessment, suggesting that remote work environments may be fostering more inclusive communication. However, the exclusive evaluation of PWD as Neutral (9.5\%) in the Likert scale highlights unique challenges faced by this group, which appear to be invisible to leaders and teammates.
% This divergence underscores the importance of not only implementing inclusive communication strategies but also continuously monitoring and adapting them to address the specific needs of PWD.

\subsubsection{Coordination Results}
\begin{figure}[h]
  \centering
  \includegraphics[width=\linewidth]{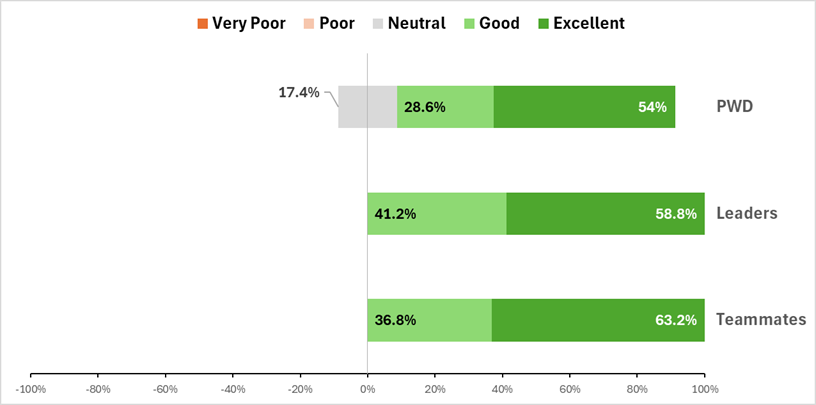}
  \caption{Evaluation of the \textbf{coordination} of activities}
  \Description{Horizontal bar chart showing how three groups—Persons With Disabilities (PWD), Leaders, and Teammates—evaluate coordination of activities in remote teams. Each group’s bar is segmented into five categories: Very Poor (dark orange), Poor (light orange), Neutral (gray), Good (light green), and Excellent (dark green). Percentages are displayed for the segments.}
  \label{fig:coordination}
\end{figure}

Remote work is also seen as facilitating transparency and ease of coordination (as illustrated in the second bar in Figure \ref{fig:coordination}), as all activities and communications are documented and can be revisited if needed. 

Among PWD, 17.4\% rated coordination as "Neutral", 28.6\% as Good and, 54\% as "Excellent" Leaders showed only positive view, with 41.2\% selecting "Good" and 58.8\% choosing "Excellent". Teammates expressed the highest level of satisfaction, with 36.8\% rating coordination as "Good" and 63.2\% as "Excellent".

Most of the repondents rated coordination activities  as ``Excellent''~\ref{fig:coordination}). However, when comparing the evaluation of communication, we can notice a  decrease of Excellent evaluations and rise of Good and Neutral perceptions, specially by PWD, which remains the only group rating in Neutral range. This distribution highlights that a significant portion of PWDs experience challenges that are not perceived or acknowledged by others. These challenges may include difficulties in tracking task assignments, understanding dependencies, or receiving timely updates on workflow transitions. 

%Most of the teammates rated activities coordination as ``Excellent'', followed by Leaders and PWD (\ref{fig:coordination}).

\subsubsection{Cooperation Results}
\begin{figure}[h]
  \centering
  \includegraphics[width=\linewidth]{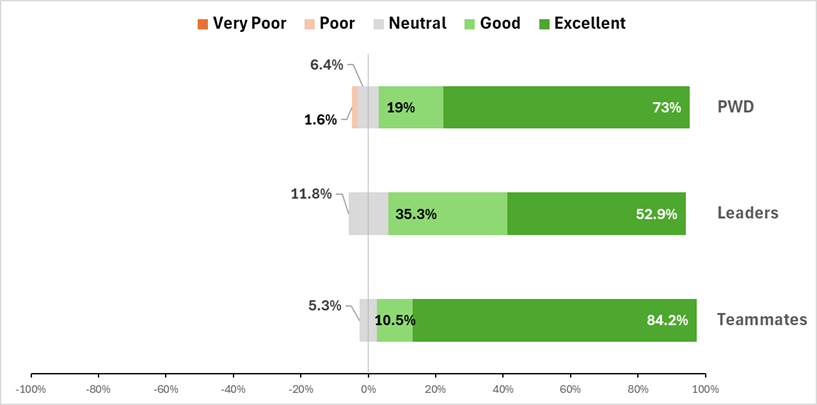}
  \caption{Evaluation of the remote \textbf{cooperation}}
  \Description{Horizontal bar chart showing how three groups—Persons With Disabilities (PWD), Leaders, and Teammates—rate cooperation in remote teams. Each group’s bar is segmented by five satisfaction levels: Very Poor (dark orange), Poor (light orange), Neutral (gray), Good (ligt green), and Excellent (dark green). Percentages are displayed for the results.}
  \label{fig:cooperation}
\end{figure}

The high level of positive (Excellent and Good) evaluations about cooperation across all groups (See Figure \ref{fig:cooperation}) suggests that respondents generally feel successful in collaborative efforts, even in this remote context. 
For PWD, 1.6\% evaluates as Poor, 6.4\% as Neutral, 19\% Good and 73\% Excellent. For Leaders, 11.8\% Neutral, 35.3\% Good and 52.9\% Excellent. For Teammates, data shows  5.3\% Neutral, 10.5\% Good and 84.2\% Excellent.
Analyzing data one can notice that a small percentage of PWD reported poor cooperation, while Teammates and Leaders did not report any negative categories. Again, this finding underscores that, despite overall positive perceptions, there remain specific barriers or unmet needs for some PWD in the context of remote cooperation. The wider variation of Leaders evaluation among Excellent and Neutral, may indicate that this group can perceive, better than teammates, the cooperation challenges in mixed-ability development teams.

% \textcolor{red}{acho que pode retirar este texto pois ele me parece dizer a mesma coisa que os paragrafos anteriores}All three investigated groups reported predominantly positive experiences with remote cooperation, with Teammates being the most satisfied, followed by PWD, and then Leaders (Figure~\ref{fig:cooperation}). The presence of Neutral and a few negative responses, especially among PWD and Leaders, underscores the importance of ongoing efforts to identify and address the specific needs and challenges of people with disabilities in remote work settings.

\subsection{Challenges}

\subsubsection{Leader's Perspective}
Leaders expressed in qualitative data some knowledge and concern about accessibility in interactions with PWD that they lead.
\begin{itemize}
    \item \textbf{Recognition of Individual Needs}: Some leaders acknowledge the need for tailored approaches or accommodations, especially for specific disabilities, but generally feel these are manageable.
    \item \textbf{Accessibility Limitations in Tools:} Leaders recognize that some digital tools used in remote work are not fully accessible, particularly for visually impaired team members.
    \item \textbf{Communication Nuances and Preparation}: Some leaders note that effective communication with PWD may require extra preparation, adaptation, or the use of interpreters, especially for those with hearing or visual impairments.
    \item \textbf{Limited Awareness of Disability}: There is sometimes a lack of awareness or understanding of the specific challenges faced by PWD, as some leaders are not even aware of the disability status of their team members.

\end{itemize}

\subsubsection{Teammates' Perspective}
Non-disabled colleagues generally perceive remote mixed-ability teams as inclusive, communicative, and collaborative, with few explicit barriers. However, they do recognize some technical and accessibility challenges, particularly regarding communication tools and integration of new team members. 

\begin{itemize}
    \item \textbf{Accessibility Limitations in Digital Tools}: There is a perception that some digital platforms or tools used in remote work are not fully accessible, particularly for visually impaired team members, which can hinder effective participation.
    \item \textbf{Technical Communication Barriers}: Some colleagues report that technical communication can be more challenging when interacting with certain groups of PWD, especially when the tools or formats are not fully accessible or when the subject matter is complex.
\end{itemize}

Many team members do not perceive any difference in treatment, participation, or performance between PWD and non-disabled. It would be a positive aspect, however, when comparing with PWD's Perspectives one could evaluate as a misperception of PWD unique charactheristics and needs.

\subsubsection{PWD's Perspective}
Despite the positive evaluation about remote-first and hybrid work satisfaction, PWD still faces barriers in accessibility, communication, and coordination.
\begin{itemize}
    \item \textbf{Accessibility Barriers in Tools and Processes}: Difficulties arise when digital tools, platforms, or documentation are not fully accessible, especially for visually impaired or neurodivergent team members. This includes issues with screen readers, lack of alternative formats, or inaccessible collaborative features.
    \item Communication Misalignments and Overload: Challenges in communication include unclear or fragmented information, preference mismatches (e.g., written vs. video), and difficulties in synchronous meetings, especially for those with neurodivergence or hearing impairments.
    \item \textbf{Coordination and Task Clarity Issues}: Some PWD report confusion about task assignments, lack of clarity in transitions between tasks, or insufficient documentation of decisions, which can be exacerbated in remote settings.
    \item \textbf{Social Isolation and Overload}: Remote work can intensify feelings of isolation, especially when socialization is limited or when there is pressure to be constantly available, leading to overload or anxiety.
\end{itemize}

\subsection{Positive Experiences}
\subsubsection{Leader's Perspective}
Under the leaders' perspective, some initiatives contribute to PWD's positive experience in software development remote teams:
\begin{itemize}
    \item \textbf{Intentional Inclusion and Relationship Building}: Leaders who intentionally foster connection, create opportunities for feedback, and seek to understand the individual beyond the professional role contribute to a more inclusive environment.
    \item \textbf{Structured and Transparent Processes}: The use of well-defined agile methodologies, clear documentation, and transparent communication channels supports the inclusion of PWD in remote teams.
    \item \textbf{Collaborative and Supportive Team Culture}: A culture of collaboration, mutual support, and shared decision-making is seen as a key factor in positive experiences for all team members, including PWD.
% Finally, Leaders also highlight the proactive and engaged participation of people with disabilities, noting that their contributions are on par with those of non-disabled peers.
\end{itemize}
In summary, for Leaders, the overall experience is positive, with remote work supporting both individual and collective success, fostering a collaborative environment, and enabling inclusive practices.    

\subsubsection{Teammate's Perspective}
Positive experiences are fostered by proactive support, open communication, and flexible task coordination, though there is room for greater awareness and adaptation to the specific needs of PWD.
\begin{itemize}
    \item \textbf{Proactive Team Support and Willingness to Help}: Team members actively support each other, are willing to help, and adapt communication or processes as needed to ensure everyone can contribute.
    \item \textbf{Clear and Open Communication Channels}: The use of clear, direct, and open communication channels (including chat, video, and interpreters) is seen as essential for effective collaboration and inclusion.
    \item \textbf{Positive Attitudes and Engagement}: Recognition of the proactive, engaged, and committed attitudes of colleagues with disabilities, which contributes to a positive team dynamic.
\end{itemize}

\subsubsection{PWD's Perspective}
PWD in remote, mixed-ability software development teams experience both unique challenges and significant benefits. Barriers are often mitigated by supportive team cultures, clear communication, flexible arrangements, and the inherent advantages of remote work. 
\begin{itemize}
    \item Supportive and Collaborative Team Culture: Many PWD highlight the importance of a collaborative, respectful, and inclusive team environment, where colleagues are proactive in offering help and adapting to different needs.
    \item \textbf{Clear and Respectful Communication}: Teams that prioritize clear, objective, and respectful communication—using multiple channels and adapting to individual preferences—are seen as highly inclusive.
    \item \textbf{Flexibility and Empathy in Work Arrangements}: Flexibility in communication modes, meeting participation, and task management, combined with empathy from colleagues and leaders, greatly enhances the remote work experience for PWD.
    \item \textbf{Personal and Professional Benefits of Remote Work}: Remote work is valued for improving quality of life, saving commuting time, and allowing for better health and family time, which is especially significant for PWD.
\end{itemize}

\section{Discussion}

Our results show that remote work is broadly perceived as a positive modality by people with disabilities, their teammates, and leaders in mixed-ability software development teams. This finding resonates with recent large-scale studies reporting that remote-first arrangements offer flexibility, autonomy, and improved work-life balance, which are especially valued by software engineers~\cite{ford2021tale}. For PWD in particular, the benefits extend beyond convenience: remote work reduces the physical and logistical barriers associated with commuting and provides a more adaptable environment for managing health, accessibility needs, and personal responsibilities. These findings reinforce earlier work suggesting that remote settings can mitigate structural barriers for marginalized groups in software engineering~\cite{FurtadoCTP21}. 

Despite the overall positive perception, our study highlights that PWD continue to experience unique challenges in communication, coordination, and cooperation—dimensions critical for distributed software engineering \cite{Herbsleb:FOSE:2007}. For example, while teammates and leaders largely rated communication as excellent, only PWD reported neutral or negative experiences. This discrepancy suggests that accessibility barriers may remain invisible to non-disabled peers, echoing prior research showing that distributed work can amplify subtle inequities in information sharing and awareness~\cite{ford2021tale}. Similarly, while coordination was positively assessed, PWD reported more difficulties with task clarity and workflow transitions, pointing to persistent challenges in ensuring transparency and equal access to team knowledge. These issues underline the importance of accessible documentation practices and the adoption of inclusive coordination tools that integrate well with assistive technologies.

The consistently high evaluations of cooperation across groups suggest that collaborative practices in remote teams may support inclusion when combined with a culture of mutual support and respect. Yet, the fact that a small proportion of PWD still reported negative experiences underscores that cooperation does not automatically guarantee inclusion. Leaders, in particular, showed a wider spread of responses, suggesting that they may be more attuned to challenges in mixed-ability teams, but also less consistent in implementing inclusive practices.  Finally, our findings suggest that positive experiences are strongly linked to team culture and organizational practices. Supportive colleagues, empathetic leadership, and structured yet flexible processes appear to act as enablers of inclusion. These results are consistent with prior literature highlighting that technical solutions alone are insufficient to guarantee accessibility; instead, a combination of technological, organizational, and cultural factors is needed~\cite{GraziotinFWA18}.

\section{Limitations}

Our study has several limitations that should be considered when interpreting the findings. \textit{External validity.} Data were collected within a single large Brazilian consulting company, which may limit generalizability to organizations of different sizes, cultures, geographies, or industry domains. The organizational policies, tooling ecosystem, and inclusion programs in this setting may not reflect those of other companies. Moreover, the survey and interviews were conducted in Portuguese and analyzed in English for reporting; despite careful translation, subtle meanings may have been lost or altered. \textit{Sampling and nonresponse bias.} Participation was voluntary and disseminated through internal channels, which may have favored respondents with stronger opinions or higher engagement with inclusion topics. The distribution across roles was imbalanced (e.g., fewer leaders than PWD or teammates), and interviews focused only on PWD; thus, perspectives from non-disabled teammates and leaders were not qualitatively explored with equal depth. \textit{Measurement and construct validity.} The survey relied on self-report measures, including self-identification of disability type and single-item Likert evaluations for key constructs (e.g., satisfaction, communication quality), which can introduce common-method bias and limit construct reliability. We did not triangulate perceptions with behavioral/trace data (e.g., repository, calendar, or communication logs) or validated psychometric scales, which could strengthen inferences about collaboration quality and inclusion. \textit{Design choices.} To prevent duplicate responses, the survey was not anonymous. While this improves data integrity, it may have inhibited disclosure of negative experiences or sensitive information, especially in a workplace context. \textit{Qualitative analysis.} Although we followed a rigorous thematic analysis with iterative codebook refinement and author consensus, qualitative coding inherently involves researcher interpretation; we did not conduct member checking with participants or an external audit to further validate themes. \textit{Temporal scope.} The study is cross-sectional (survey open in August 2025 and interviews in the same period), capturing a snapshot rather than longitudinal dynamics as policies, tools, and team compositions evolve. \textit{Accessibility of instruments.} Despite piloting and accessibility reviews, some questions or survey interfaces may still have posed barriers (e.g., screen reader nuances), potentially affecting response completeness or depth. Finally, our focus on software developers (in interviews) means findings may not fully capture challenges and enablers experienced by other roles in software teams (e.g., QA, UX, PM). These limitations suggest caution in generalizing the results and motivate the multi-organization, longitudinal, mixed-method extensions we outline in future work.

\section{Implications}

Our findings have several important implications for software companies seeking to foster more inclusive and effective remote work environments. First, organizations should recognize that while remote work provides flexibility and autonomy for people with disabilities (PWD), it also introduces specific accessibility challenges. Companies need to invest in accessible collaboration tools, ensure compatibility with assistive technologies, and provide training for both leaders and teammates on inclusive communication practices. Establishing clear policies that address disclosure, reasonable accommodations, and support for hybrid teams is essential to prevent PWD from being unintentionally excluded from key interactions. Moreover, leaders should proactively monitor and address invisible barriers, such as coordination difficulties or uneven access to shared knowledge, which may not be readily apparent to non-disabled team members. Implementing regular feedback loops, accessibility audits, and inclusive team rituals can help organizations create a culture of awareness and continuous improvement.

For researchers, this study highlights the need to further investigate the intersection of disability, accessibility, and software engineering practices in remote and hybrid contexts. Future research can build on our results by designing and evaluating interventions aimed at reducing identified barriers, such as accessible software development tools, inclusive meeting facilitation strategies, or adaptive leadership training programs. Additionally, researchers should explore cross-organizational and cross-cultural studies to understand how different environments shape the experiences of PWD in software teams. Methodologically, combining qualitative insights with behavioral or trace data could offer a more holistic view of collaboration dynamics and accessibility issues. By deepening this line of research, the software engineering community can generate actionable knowledge to guide both industry practices and public policies that support equitable participation of PWD in the technology sector.

\section{Conclusion and Future Work}

The inclusion of people with disabilities (PWD) in software development teams remains a critical yet underexplored challenge, especially in the context of remote and hybrid work. While remote work has become increasingly prevalent in the technology sector, little is known about how accessibility needs and disability-related barriers affect collaboration, communication, and coordination within mixed-ability software development teams (SDTs). In this paper, we investigated this problem through a mixed-methods study that combined an online survey with \totalSurveyResponses participants and 14 semi-structured interviews with PWD. Our study was guided by three research questions focusing on how disabled and non-disabled team members evaluate remote work, the main challenges faced by PWD, and the factors that promote positive remote work experiences. Our findings reveal that while remote work can bring significant benefits for PWD, such as flexibility and autonomy, it also introduces unique barriers related to accessibility of tools, lack of awareness among non-disabled team members, and difficulties in communication and coordination. Moreover, we identified practices and strategies that foster inclusion, such as accessible technologies, supportive leadership, and proactive team culture, providing actionable insights for organizations seeking to create more equitable and accessible remote work environments.

Building on these results, several avenues for future research emerge. First, longitudinal studies are needed to examine how the experiences of PWD evolve over time as remote work practices and organizational policies mature. Second, research could explore the design and evaluation of specific interventions, such as accessibility-focused collaboration tools or training programs for team leaders and members, to address the barriers identified in this study. Third, expanding the scope beyond a single organization and geographic context would provide a broader understanding of the challenges and enablers of inclusion in diverse settings. Finally, integrating perspectives from other stakeholders, such as human resources professionals, product managers, and accessibility experts, could offer a more comprehensive view of how to foster sustainable inclusion for PWD in software engineering. Through these future efforts, the software engineering research community can continue to advance knowledge and practice toward building truly inclusive, accessible, and high-performing teams.

%\section{Acknowledgments}
%já tá correto esse texto

%We would like to thank Conselho Nacional de Desenvolvimento Científico e Tecnológico (CNPq) - grant numbers 420406/2023-9 and 442779/2023-2 - for partially funding this study, and to Coordenação de Aperfeiçoamento de Pessoal de Nível Superior (CAPES) - Finance Code 001 - for providing scholarship funding to the first, second, and fifth authors. We thank GitHub Copilot (Microsoft) for the empirical study and Grammarly for the review of our writing.

\bibliographystyle{ACM-Reference-Format}
\bibliography{references}

\end{document}